\documentclass[fleqn,usenatbib]{mnras}
\pdfoutput=1
\usepackage{newtxtext,newtxmath}

\usepackage[T1]{fontenc}

\DeclareRobustCommand{\VAN}[3]{#2}
\let\VANthebibliography\thebibliography
\def\thebibliography{\DeclareRobustCommand{\VAN}[3]{##3}\VANthebibliography}

\usepackage{graphicx}	
\usepackage{amsmath}	

\title[Automated coronal width detection]{Automated detection and analysis of coronal
active region structures across Solar Cycle 24}

\author[D. G. Gass et al.]{
Daniel G. Gass,$^{1}$\thanks{E-mail: DGGass@uclan.ac.uk}
R. W. Walsh,$^{2}$\thanks{E-mail: RWWalsh@uclan.ac.uk}
\\
$^{1}$Jeremiah Horrocks Institute, School of Natural Sciences, University of Central Lancashire, Preston, Lancashire, Fylde Rd PR1 2HE, UK\\}

\date{Accepted XXX. Received YYY; in original form ZZZ}

\pubyear{2023}

\begin{document}
\label{firstpage}
\pagerange{\pageref{firstpage}--\pageref{lastpage}}
\maketitle

\begin{abstract}
Observations from NASA’s Solar Dynamic Observatory Atmospheric Imaging Assembly were employed to investigate targeted physical properties of coronal active region structures across the entirety of Solar Cycle 24 (dates). This is the largest consistent study to date which analyses emergent trends in structural width, location, and occurrence rate by performing an automatic and long-term examination of observable coronal limb features within equatorial active region belts across four extreme ultraviolet wavelengths (171, 193, 211, and 304 angstroms). This has resulted in over thirty thousand observed coronal structures and hence allows for the production of spatial and temporal distributions focused upon the rise, peak and decay activity phases of Solar Cycle 24. Employing a self- organized-criticality approach as a descriptor of coronal structure formation, power law slopes of structural widths versus frequency are determined, ranging from -1.6 to -3.3  with variations of up to 0.7 found between differing periods of the solar cycle, compared to a predicted Fractal Diffusive Self Organized Criticality (FD-SOC) value of -1.5. The North-South hemispheric asymmetry of these structures was also examined with the northern hemisphere exhibiting activity that is peaking earlier and decaying slower than the southern hemisphere, with a characteristic "butterfly" pattern of coronal structures detected. This represents the first survey of coronal structures performed across an entire solar cycle, demonstrating new techniques available to examine the composition of the corona by latitude in varying wavelengths at selected altitudes. 

\end{abstract}







\section{Introduction}
\subsection{The Solar Cycle and Active Regions}
The approximately eleven year solar activity cycle as part of the larger twenty two year magnetic cycle are one of the clearest observable indicators of the Sun's complex and dynamic magnetic field, with consequences for the wider heliosphere and space weather. However, the specifics of how varying magnetic activity can propagate outwardly into the Corona, is still not fully understood. 

Coronal structures such as loops and plumes are observable extensions of the sun's magnetic field \citep{2014LRSP...11....4R}. Though these structures have been extensively imaged over Solar Cycle 24 by instruments such as NASA's Solar Dynamics Observatory (SDO) Atmospheric Imaging Assembly (AIA) \cite{2012AIA} \cite{2012SDO}. These rich and expansive datasets have not always been fully leveraged to take advantage of this long term coverage. Previous studies have focused on aspects of specific properties of coronal structures such as the cross sectional profiles and morphology \cite{2020Klim}, lengths \cite{2018ApJ...868..116D}, intensities and temperatures\cite{2017ApJPlasmaTemp}, and the interaction of complex magnetic regions \cite{2015ApJ...815....8R}, however these are typically limited in scope and only utilize small portions of the data available, such as specific active regions and time periods. The evolution of these properties and any possible connection across the solar cycle is not fully appreciated. 

Equatorial active regions are areas of predominantly closed magnetic flux occurring roughly thirty degrees above and below the line of the solar equator (\cite{2015AR}). These are referred to as the active region belts and are indicators of the phases of magnetic activity across the solar cycle \citep{2015CycleReview}. The solar magnetic field can vary in strength drastically by region and altitude and is also associated with a high degree of dynamism and magnetic field intensity variation \citep{2021Coronalfield,2009ARstrength}, with high temperature and velocity variations within plasmas and structures. This makes it a particularly compelling region for research into the interactions of plasma with these strong, varying magnetic fields \cite{2021CoronaDyn1,2017CoronaDyn2}.

These regions of magnetic activity can exhibit highly complex braiding behaviours within magnetic flux tubes on a wide variety of spatial scales, which may contribute to coronal heating (\cite{2022CoronalBraiding}). A combination of motions from photospheric Alfvén waves (AC Heating, \cite{1997ACHeating}) to the magnetic tension causing local reconnections and realignments of the magnetic field and dissipation between current sheets (DC nanoflare \cite{2004NanoflareHeating} and Joule heating \cite{2019JouleHeating}, is likely contribute to the heating of coronal plasma. As heated plasma fills the flux tubes which contain them, this plasma radiates at specific wavelengths, and observable coronal structures are formed. The placement and observed width of these structures may therefore be informative of the physical processes and the magnetic conditions which produced them.

\subsection{Coronal Structures and Observation}
Coronal structures are pervasive fundamental features of the solar corona \citep{2014LRSP...11....4R}. These structures vary greatly in length, from a few hundred kilometers in very small loops to several solar radii for extended plumes, formed from plasma heated by magnetic reconnection and supported by magnetic flux tubes emerging from the solar interior. The extremely high temperatures and low pressures that typify the corona have been subject to extensive study by solar astrophysicists. There are a variety of models utilizing a wide range of MHD (Magnetohydrodynamical) based heating scenarios (such as DC filament \citep{1984DCheating} and AC Wave Heating \citep{1997ACHeating}), though more limited is extensive observational data to compare these models to. 

Limitations of coronal observations are imposed by the instrumental effects, and of viewing conditions of the corona. Some of these can be mitigated by choosing structures visible at the coronal limb, but others - such as the point spread function and charge spreading effects are nontrivial to deconvolve from images\cite{2013ApJPSF} and can effectively destroy information of smaller structures.

A summary on existing coronal loop observations made with various instruments is provided by \cite{2017Aschwanden}; they outline the decreasing minimum observable loop width that is brought about as a consequence of improving instrumental resolving power.  \cite{2017Aschwanden} describes an observed loop width as a combination of the true width size and any possible perceived loop broadening which are observational effects, described as the following;

\begin{equation}
    w^2_{obs} = w^2_{psf} + w^2_{true} + w^2_{noise}
    \label{eq:wspreading}
\end{equation}

where \textit{w} = observed loop width, \textit{w\textsubscript{true}} = the “true” loop width, or how large the structure would appear if viewed without any limiting effects, \textit{w\textsubscript{psf}}= the point spread function caused by the “spreading” of charge in an electronic CCD when a photon is absorbed by a pixel in the detector, and \textit{w\textsubscript{noise}} = broadening caused by noise effects such as Poissonian photon noise, dark current, readout noise, etc.  For EUV AIA wavelengths, this is approximately 2.7 pixels.

\subsection{Coronal Widths and Self Organized Criticality}

Despite the above, statistical methods such as S.O.C modelling (Self Organized Criticality \cite{1988PhRvA..38..364B}) can be useful in analysing the behaviour of nonlinear, size invariant systems. The mechanism for DC heating, which is suspected to be a major cause contributing to the processes of reconnection which form visible coronal structures is potentially describable as an S.O.C process. That is, that the requirements of stochastic addition (energy deposited randomly to subresolved filaments within the footpoints of coronal structures), the existence of a local critical threshold (the amount of energy which is needed to trigger the process of magnetic process), the multi-scale nature of the space (subresolved flux tubes of scale lengths of potentially as few as sub ten km in a space up to 10,000s km), and the presence of a global critical state by which small scale events can spread and cascade into larger events which spread throughout the space (local magnetic reconnection perturbs adjacent filaments and causes further reconnection events to occur throughout the footpoint of a coronal structure) mean that some characteristic statistics of S.O.C processes might be detectable within large enough samples of these coronal structures. 

A strength of the S.O.C approach is that it does not rely on precise understanding of the exact mechanisms which are operating within the observed region. Highly complex reconnection reconstruction which would require complex 3-D MHD simulation  are not required for approximating the contributions of a wide variety of effects and local conditions as a local requirement of exceeding some critical threshold in some  parameter(s), alleviating the requirement for the entire system to be solved analytically for predictions of observational parameters. S.O.C has been shown to apply to a large number of physical and astrophysical phenomenon, \cite{2016SOCConcepts}, including in reconnection in solar plasmas (flares, CMEs, solar energetic particles etc..) \cite{2016SOCPlasma} and in coronal loop widths themselves \cite{2017Aschwanden}.  Although S.O.C based analysis will allow for a greater degree of physical interpretation of coronal loop populations and their connection to the magnetic changes which occur in the solar cycle in bulk, there are some caveats it must be noted that the aforementioned observational limitations will introduce a threshold cutoff effect to the power law of observed coronal widths. This effect thresholds the distribution of observed S.O.C events viewed by the AIA, including coronal structures below the 2.7 pixel limit described above - and alters their measured power laws. These observational effects combined with alterations to underlying physical and magnetic processes driven by the solar cycle could cause a measurable deviation from a predicted S.O.C gradient and by examining their relation to ideal values derived from stochastic, fractal events in n-dimensional S.O.C space (1.5 in the case of coronal loop widths, which are taken to be analogues for the size s of cascading events). 

The power law predicted by the ideal Fractal Diffusive (FD) S.O.C is indicative of the presence of coupled, driven oscillators and the dissipative non-linear avalanching behaviour, which spreads across all available scale lengths given a stochastic driver. As it can be safely assumed that the simple S.O.C approach is not wholly descriptive of the physical phenomenon occurring in the environment of the solar corona during loop formation, deviations from this ideal value can be described as combinations of observational effects, and specific variations as described above. These variations can also be described by modification of the simple ideal case.

In the context of this work, if coronal structures do follow S.O.C like distributions, then a collection of their physical widths should be describable by a power law. By measuring a wide variety of coronal structures in varying conditions, a profile can be developed and a power law fitted and measured.

Prior results from studies of active regions and demonstrate S.O.C distributions within coronal loop populations found that power law slopes of widths from these coronal structures range from roughly 2.7 - 3.3  for active regions viewed in multiple AIA filters and around 1.39 in Hi-C populations. \citep{2019Zhiming,2017Aschwanden} This compares to an "ideal" FD-SOC gradient of 1.5

Power law slopes for varying wavelengths at various points in the solar cycle can be compared and measured, and the exact difference between $\alpha$\textsubscript{w} in various stages of the solar cycle and with $\alpha$\textsubscript{s} can be quantified and analysed. The closeness of fit of these observed profiles to the previously mentioned power-law will be used to analyse loop widths, comparing different populations of loop widths from various points across the entire recent solar cycle in the first instance. These power laws can form a basis for analysis of coronal structure populations across varying time periods and wavelengths.

S.O.C then might be useful as a type of statistical probe of emergent behaviours in the corona. As it takes no assumptions or is informed by any physics within the region of loop formation, it cannot directly provide information as to physical conditions or which specific mechanism or threshold condition is required to produce the observed collection of structures.


\subsection{North South Hemispheric Asymmetry}

Additionally, a long term approach to examining coronal structures could analyse north vs south hemispheric asymmetry. Imbalance in northern and southern solar magnetic activity has been observed for several decades \citep{bell1958some} and has been explored extensively in a variety of solar phenomenon such as; sunspot activity \citep{2021Ap&SS.366...16J}, interplanetary energetic and geomagnetic indicators \citep{2012Ap.....55..127E}, sunspot rotation rates \citep{2018ApJ...855...84X} among others. As of yet, no large scale study of asymmetry within coronal structures has been made. Though expected to be strongly related to sunspots by their shared emerging magnetic flux tubes, it is difficult for models to definitively determine the relationship between simple emerging flux tubes present at the footpoints of coronal structures and more developed observable structures higher in the solar atmosphere. This is due to complex forces such as changing plasma beta with height and mechanical warping/tension of these structures. By examining this asymmetry in context with the different plasma temperatures of varying EUV bands, greater detail of plasma temperatures and their changing distributions across the solar cycle can be examined. The aim of this work then is to identify and record the largest possible population of coronal structures across solar cycle 24 by measuring and studying the evolution of coronal loop width, latitude, and occurrence frequency. This will allow for novel analysis of the corona - measuring specific changes to many of these parameters and contrasting them to other well known indicators of solar activity such as sunspots, models of predictions of coronal loop parameter distribution, and against the properties of other  loops recorded in multiple EUV wavelengths.

As well as coronal widths, the hemispheric asymmetry of structures can indicate the behaviour of the corona and it's reaction to the changing activity within the asymmetric solar dynamo. Such examinations would allow for a coronal region to be "mapped" throughout the solar cycle, and examined for trends and distributions which may be characteristic of the solar atmosphere and the magnetic field as it moves into the outer solar atmosphere. Such examination may prove useful for determining the degree of similarity between the placement coronal structures and their corresponding photospheric footpoints, such as active region sunspots, as well as identifying divergences from these expected positions in different temperature regimes. The behaviour of magnetic flux tubes and how they change as a consequence of emergence and height above the photosphere is not completely understood (\cite{2018Fluxtubes}). The mechanisms of how  these mechanisms have for space weather and the interaction of these solar magnetic fields with the outer solar atmosphere, this could represent an alternative way to approach studying the shape and the dynamics of the coronal atmosphere throughout a solar cycle.

In the following section, the methodology of an automated approach to identification and measurement of coronal loops in AIA EUV images is outlined, and the techniques of analysis performed upon them described.

\section{Method}
To analyse the properties of structures of interest within the corona, they must first be identified and traced; in the case of coronal loop structures, this can be challenging. Difficulties of variable geometries owing to the relative angle of the observed structure to instrument mean that geometry and width determination must be considered. This is not always feasible to perform in all portions of the corona in an automatic fashion, as there are a number of assumptions which must be made about the underlying loop orientation and background. This involves complex magnetic and structural topologies such as by visually overlapping structures along the line of sight\citep{2004CoronaTopology}, which can introduce errors to subsequent measurement \citep{2013ApJ...773...94M}. 

Additionally, the contribution of background/foreground emission along the integrated line of sight can be significantly greater than the emission of coronal loop structures, and attempts to properly isolate loop intensity from background measurements can be challenging. \citep{2014LRSP...11....4R}

To mitigate these issues, coronal loops at the limb were chosen as the basis of study for measurement and analysis. Limb loops are by definition observed away from the bright solar disc, thus simplifying issues relating to contaminating background and foreground emission. This allows for measurements of loop width and intensity to be undertaken with minimal background subtraction. 


\subsection{Defining time periods across the solar cycle}

\begin{figure*}
    \centering
    \resizebox{\hsize}{!}{\includegraphics[clip]{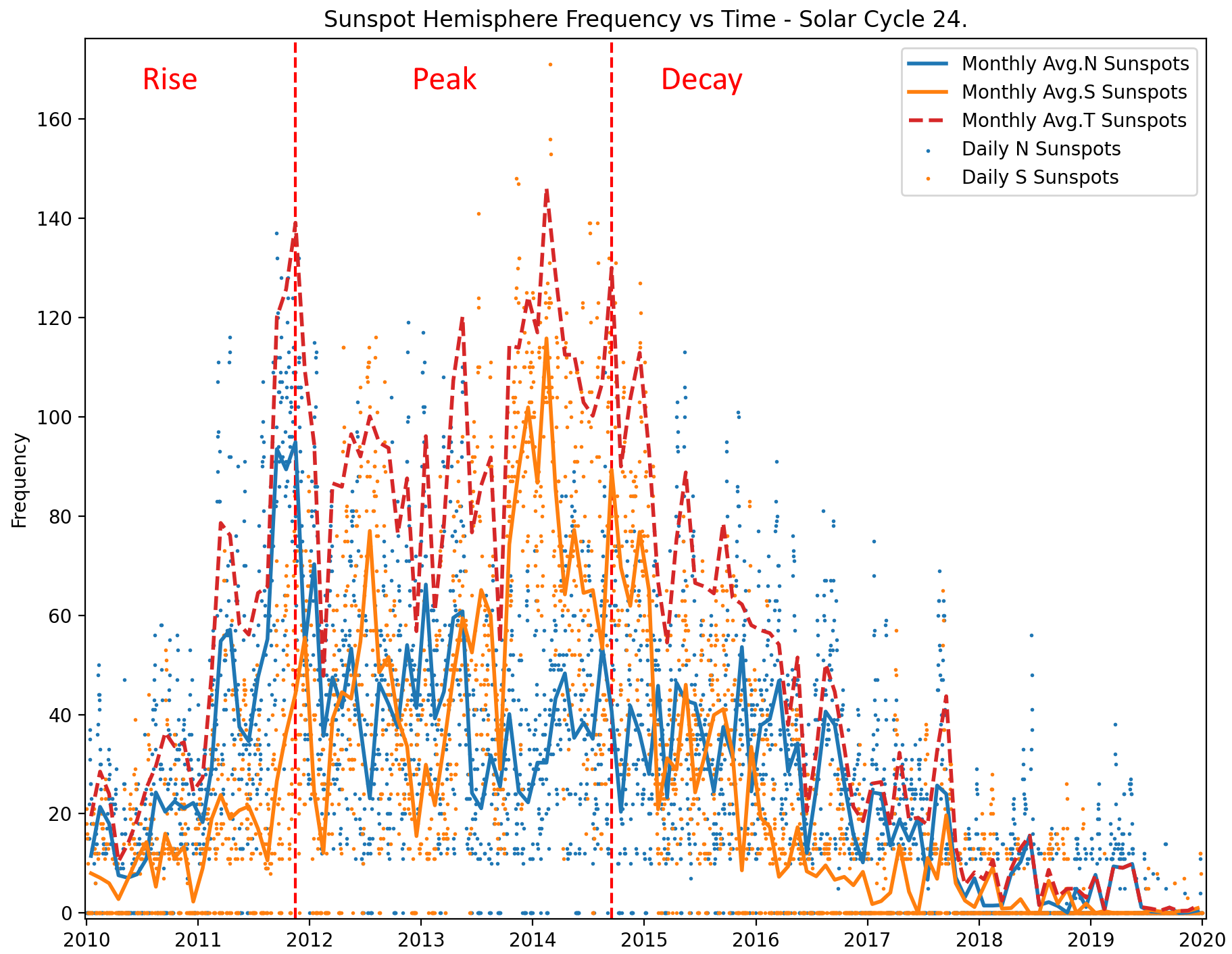}}
    \caption{Sunspot hemispheric activity (y axis) vs time (x axis). Blue indicates northern activity, orange indicates southern activity. Dots and lines indicates count by day and monthly average respectively.}
    \label{fig:NSspots}
\end{figure*}

Solar Cycle 24 is observed to be from 2010 to 2020. To examine the overall cycle, it's duration was broken down into three main time periods as outlined in Figure \ref{fig:NSspots} and in more detail below;

\begin{itemize}
\item {The \textbf{"Rise"} phase is defined as the time of rising solar activity, starting from a base level of minimal activity  and extending to the time when the first peak of solar sunspot activity for cycle 24 is recorded. The Rise period is relatively short, beginning 13/05/2010 (the earliest captured data within the AIA database) and extending to the end of 15/11/2011.}
\item{The \textbf{"Peak"} phase is defined to be the period of maximum solar activity from 15/11/2011, where sunspot count is slowly but consistently increasing, until 15/11/2014, after which activity begins to decline consistently. This period contains the solar maximum for cycle 24, and corresponds to he highest quanity of observed sunspots within the cycle.} 
\item{Finally, the \textbf{"Decay"} period is defined to be the long period of decline in solar activity, from 15/11/2014 and extending to the end of 2019. This is the longest phase of solar activity by time, but contains a number of years of low solar activity.}
\end{itemize}
In subsequent sections, the solar cycle will be discussed by reference to the three phases mentioned above, and coronal structure populations analysed within each phase.

\subsection{North-South Hemispheric Asymmetry}
The North-South Hemispheric Asymmetry is a phenomenon by which observable measures of solar activity vary between north and south hemispheres depending on the polarity of the solar cycle. Quantitative measurement of this imbalance has been helped by the creation of N-S activity indexes\citep{articleballester1993, 1993AsymmetryIndex} as 

\begin{equation}
    AS = \frac{N - S}{N + S}
    \label{eq:NSind}
\end{equation}

Here the activity index AS is defined as the quotient of the difference between activity in the northern hemisphere N and activity in the southern hemisphere S and the total amount of activity in both hemispheres. Though this was originally applied to the context of sunspot activity, further examinations of N-S asymmetry have been performed for other indices such as solar wind speed\citep{2008IndexWindspeed}, solar flares \citep{2015Flareindices}, atmospheric solar plasma density \cite{2017plasmadensity}. Of these, the sunspot activity index is the most well known and documented indicator of solar magnetic activity, and will be used as a basis of comparison for future analysis of coronal structure activity.

Measuring and analysing N-S coronal asymmetry indexes as they change throughout time and as they vary from wavelength to wavelength may be beneficial for creating a more nuanced understanding of North-South asymmetry in coronal magnetic fields themselves.




\subsection{Image enhancement and MGN processing}

The Atmospheric Imaging Assembly (AIA) instrument \citep{2012AIA} of the Solar Dynamics Observatory (SDO; \cite{2012SDO}) contains multiple filters in the EUV regime, sensitive to coronal plasma. Filters of particular interest to this study are 171 Å, 193 Å, 211 Å, and 304 Å as they possess relatively narrow wavelength response functions and singly peaking temperature response functions. This data is reduced to level 1.5, where the image is corrected for pointing and degradation of the CCD over time. This image is then used in subsequent image enhancement techniques outlined below.

As previously mentioned, identifying coronal loops above background noise can be challenging. "Background" noise in AIA images is a combination of diffuse black body radiation, charge spreading of electrons across neighbouring CCD pixel detectors, and the "dark current" inherent to the detector without any imaged source. This is a combination of difficult viewing conditions and limitations by the physics of CCD based image detectors, but the effects of this noise can be minimized with careful image enhancement techniques.



\begin{figure*}
  \resizebox{\hsize}{!}{\includegraphics{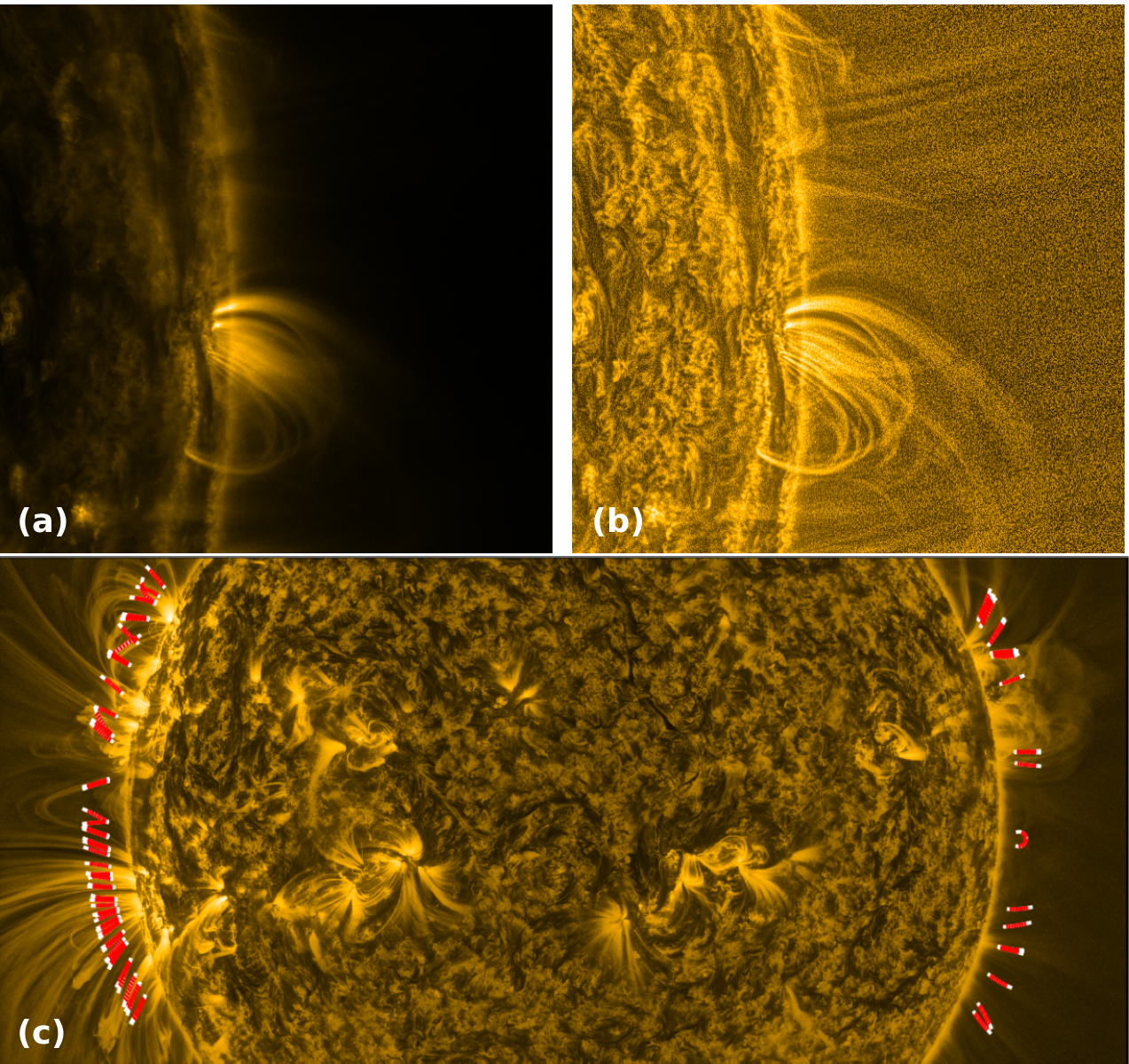}}
  \caption{A section of the coronal limb and low disc as imaged in 171 Å (left) vs MGN processed level 1.5 (right). Contrasts in coronal structure are evident, local enhancements in the diffuse limb loop structure are highlighted with their structure visually preserved to greater height above the photosphere. (c) displays the fitting of structure segments on a cropped MGN image.}
  \label{MGNExample}
\end{figure*}

Multi Gaussian Normalization (MGN), developed and outlined by \cite{2014SoPh..289.2945M}, was chosen to aid in detection of coronal structures whose widths are subsequently examined using the level 1.5 AIA image. Level 1.6 calibration was considered too computationally expensive for the entire dataset. MGN is a process by which Gaussian filters are applied to an image which is then combined with a weighted gamma transformed image, creating a composite image I. This is demonstrated in Figure \ref{MGNExample} (a) and (b), and expressed as;

\begin{equation}
    I = hC'_g +\frac{(1-h)}{n}\sum_{i=1}^{n}g_iC'_i
\end{equation}
where h is a global weighting value, C\textsubscript{g} is the global gamma transformed image, n is the number of unique Gaussian kernel widths used weighted by g\textsubscript{i} weights.

In the context of EUV observations, this technique has been used successfully in studies of Hi-C 2.1 mission data, compared to data of the same time period and region as AIA greatly enhancing the local contrast and reducing image noise (see \cite{2020Tom}, \cite{Williams_2020}).

Although MGN is used for highlighting local contrast, this can also magnify some preexisting noise. This can be alleviated by performing a time average of multiple MGN images taken in close succession. AIA usually captures images roughly twelve seconds apart (including exposure time), so although time averaging can lead to a "blurring" of structures which change position or intensity rapidly in this time window, the vast majority of coronal structures visible above the limb are stable within the span of hours, and hence should not be greatly affected. These time averaged datasets are created from images sampled sequentially (12 seconds apart), to create one composite image every 3 days (see Section 2.4); resulting in approximately 1200 composite images per wavelength across the solar cycle. MGN images are be used as a "diagnostic" layer where tracing of structures is performed, but whose coordinates are then employed to reference the corresponding pixels in the matched level 1.5 data itself. This is to avoid the effects of MGN filtering introducing nonphysical structures or altering the geometries of legitimate sources.

\subsection{Constraining image data}

Each image was cropped between 60 and 120 degrees from the solar northern pole (see Figure \ref{MGNExample}(c)), as the predominance of coronal holes and open plume-like structures increases outside of this range \citep{2020SSRv..216..117A}.



\begin{figure*}
  \resizebox{\hsize}{!}{\includegraphics{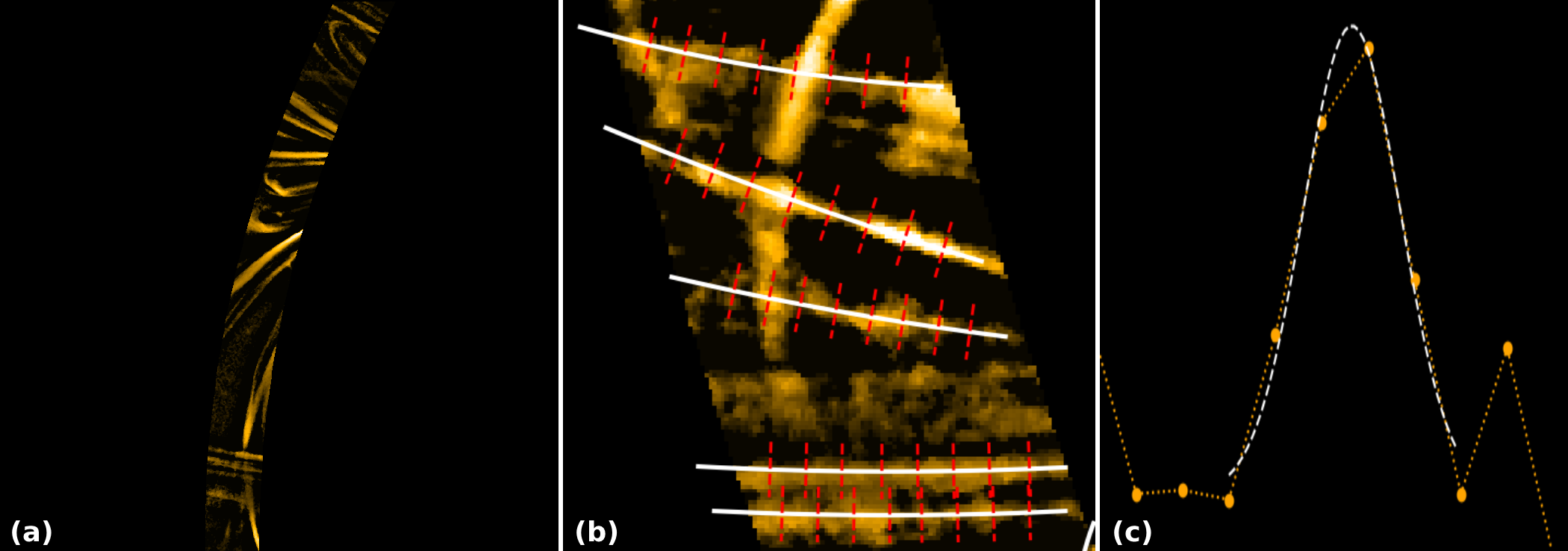}}
  \caption{A section of the coronal annulus fitted to an MGN filtered 171 A image and locally normalized and filtered to highlight coronal strands and other features. This image is used to trace geometry, and is not measured directly for structural properties.}
  \label{fig:annulusection}
\end{figure*}

Limb images are constrained to an annulus of a fixed width between 1.05 and 1.10 solar radii (roughly 35000 km). This height also allows for loops to be distinguished from their footpoints, and any emission close to the limb from visually indistinct and low lying coronal activity (such as from coronal moss and limb brightening effects). An example of this annulus fitted to a section of the limb is shown in Figure \ref{fig:annulusection} (a) and (b) .

Due to the variable lifetimes of coronal structures (which can range from hours to days or more depending on magnetic openness and degree of interaction with other nearby structures) \citep{2020ARA&A..58..441N, 2007ApJ...657.1127L}, and the variable rotation rate of structures owing to differential rotation, an optimal time cadence should be determined to prevent repeated detection of structures. To allow for this, a relationship between latitude, distance from the inner annulus and time taken to clear the annulus is found to be 

\begin{equation}
    t = \frac{\tau}{\pi} (1 - \frac{r-r_1}{2(r_2 - r_1)}) (\frac{sin^{-1}(x_1) - sin^{-1}(x_2)}{cos(\theta_c)})
    \label{eq:annulus}
\end{equation}

where $\tau$ is the rotational period at latitude $\theta_c$, r is the pixel position of the object, $r_1$, $x_1$ and $r_2$,$x_2$ is the radial and cartesian position coordinate of the inner and outer annulus at that latitude respectively. For the aforementioned annulus, this equates to a time of three days or less for the highest latitudes, and so this was chosen as the time cadence between images in the data series. 

\subsection{Width measurements}

The prerequisite for determination of power law profiles of bulk width measurements is a clear definition for "width" and a reliable method of measuring them. In this work width is defined as as the derived measurement of the distance of a gaussian profile fitted between measured minima of cross sectional brightness profiles from identified structures.



Though it should be noted that there are other possible interpretations of the morphology of coronal loops \citep{2020Klim,2022CoronalVeil}, coronal structures are typically proposed to possess roughly symmetrical cross sections. Gaussian profiles can be well fitted to coronal structures, even down to Hi-C resolutions, where it is possible to fit multiple Gaussian profiles to single cross sectional profiles due to the high pixel resolution of the instrumental data \citep{Williams_2020,2021TomStrands}.


If coronal loops are considered to possess a Gaussian cross sectional density profile, then the intensity of emission as a function of x displacement from the center of the profile can be described as 

\begin{equation}
    f(x) = a.\exp{-\frac{(x-x_p)^2}{2w^2}}
\end{equation}

where a is the height of the Gaussian profile, x\textsubscript{p} is the central position of the loop profile, x is the position away from the central position, and c is the standard deviation of the Gaussian profile.  

In many studies of coronal loops, identification and tracing is performed manually, or semi-automatically in pre-defined regions of interest. To fully utilise the extent of the AIA dataset over cycle 24, it is necessary for identification and tracking to be performed autonomously upon an image. In this case, the investigation of coronal geometry utilises custom software to process and analyse thousands of AIA EUV images in the four wavelengths mentioned previously  incorporating a modified version of the OCCULT algorithm (OCCULT \cite{2010Occult1}, OCCULT-2 \cite{2013Entrp..15.3007A}, which performs well in coronal loop detection in multiple EUV wavelengths (see appendix for links to source code). 

The annulus based approach can constrain some issues of complex loop geometry by fitting shorter, well defined loop segments, but cannot account for closed structures which re-enter the annulus at a different location. Though this is a This effect can be seen even in the use of advanced, contemporary loop tracing algorithms applied to both real and synthetic active region data (\citep{2017Aschwanden,2019Zhiming}). An example of a region of traced loops on a region of a composite MGN annulus is shown in Fig. \ref{fig:annulusection} (b).


Once a loop segment is traced across the cropped annulus, eight equidistant intensity cross sections are taken from the original level 1.5 and used to construct a mean average intensity profile. Local background is estimated by linear interpolation between local minima, and a Gaussian profile fitted to the reduced profile to estimate width. An example profile is shown in Fig \ref{fig:annulusection} (c).


\subsection{Uncertainties of Width Power Laws}

As the resulting power law gradient will be used to compare different regions, time periods, and wavelengths of coronal structural populations, the uncertainty will be derived from statistical interpretations of individual width uncertainties associated with the fitting of individual Gaussian profiles to loop cross sections. Uncertainty in loop width measurements originates in the closeness of the fit of the optimal Gaussian profile utilizing a least squares best fit algorithm. Each bin of this histogram is an equal distance in log space, with the horizontal axis being structure width in AIA pixels, and the vertical axis being occurrence frequency. By assuming that each width is normally distributed within its uncertainty, the probability that a width falls within the range of widths for a given bin can be calculated. The mean average of the probabilities for all of the widths in a bin constitute the percentage uncertainty of a particular bin, meaning the numerical error is equal to the uncertainty probability p multiplied by the number of structures in that bin N. A standard error of square root n is also applied as a 'safety'; to ensure that measurement errors better represent a process of image tracing and width extraction in which uncertainty can be difficult to quantify. The final uncertainty for each bin frequency is then represented by Equation \ref{binerror}


\begin{equation}
{\Delta}N = \bar{p}N + \sqrt{N}
\label{binerror}
\end{equation}

where $\Delta$N is the uncertainty of the frequency, ${\bar{p}}$ is the mean average probability of falling outside of a bin threshold, and N is the total frequency of coronal structures within a bin.

\subsection{Latitude Measurements}

Latitudes are recorded from 0 to 180 degrees, equivalent to the Stonyhurst heliographic coordinate system \citep{2006CoordSys}, where latitude L is equal to the coordinates $\Phi$,$\Theta$, which are horizontal and vertical displacements from the solar prime meridian and the equator respectively. In this work, all latitudes are quoted as magnitude displacements from the north pole, ie; $\Theta$ = $\pm$90.

\section{Results}

\subsection{Structural Widths}

\begin{figure*}
\resizebox{\hsize}{!}
        {\includegraphics[clip]{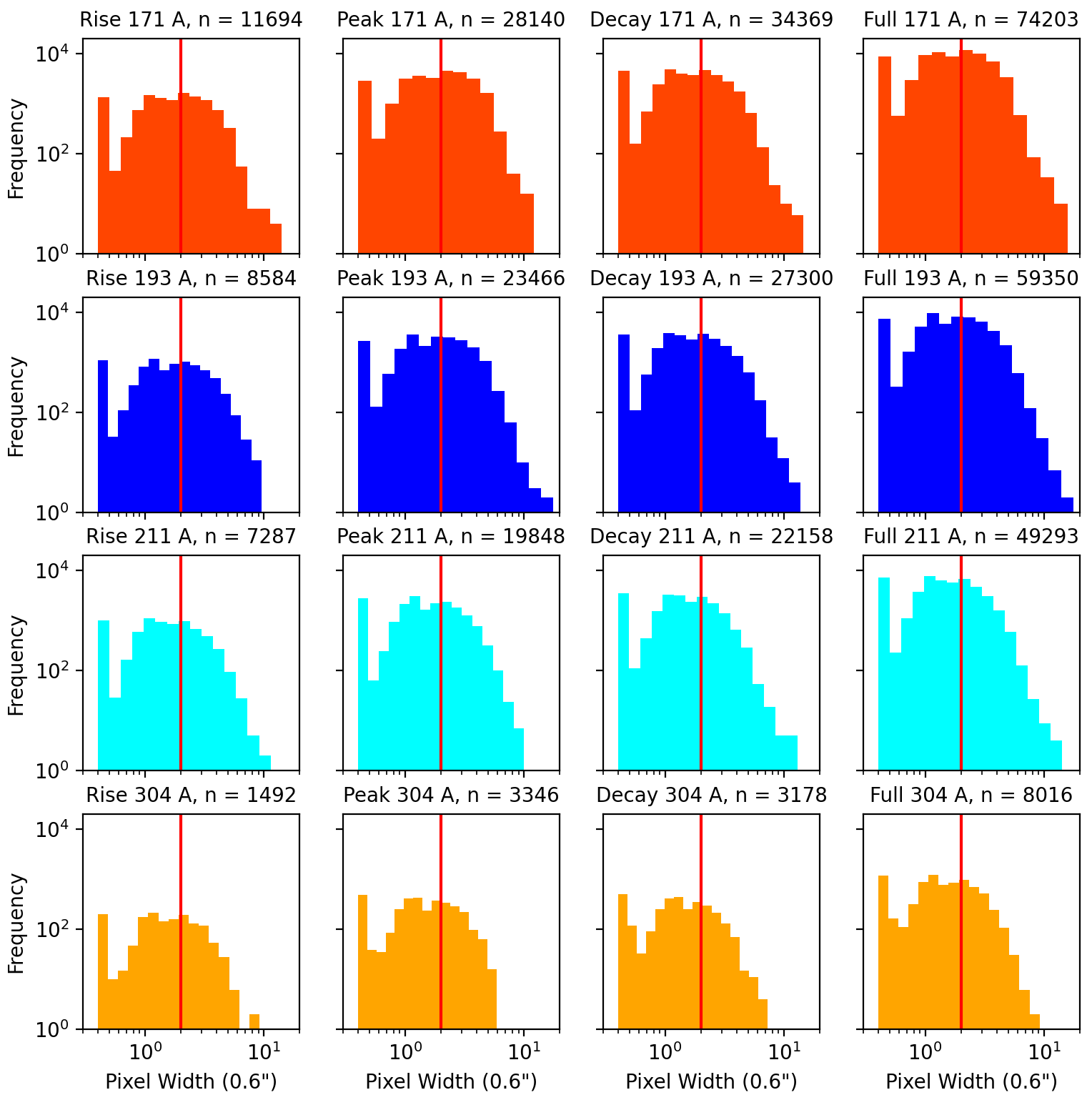}}
  \caption{Width Frequency diagram of all AIA pixel widths. The "spike" on the left hand side of each figure is an artifact of Gaussian fitting across narrow peak ranges.
          }
     \label{fig:WidthsAll}
\end{figure*}

The total number of structures detected by wavelength is demonstrated in Table \ref{table:loopsall} and Figure \ref{fig:WidthsAll}. These describe distribution of the width of structures detected in all four wavelengths throughout the solar cycle by phase and in total. These distributions include those seen below the two pixel resolution limit of the AIA instrument, and includes the prominent "spikes" between 0.3 and 0.4 pixels. This is an artifact of Gaussian fitting between narrow pixel ranges and will be ignored (see more below).

\begin{figure*}
    \centering
    \resizebox{\hsize}{!}
        {\includegraphics[clip]{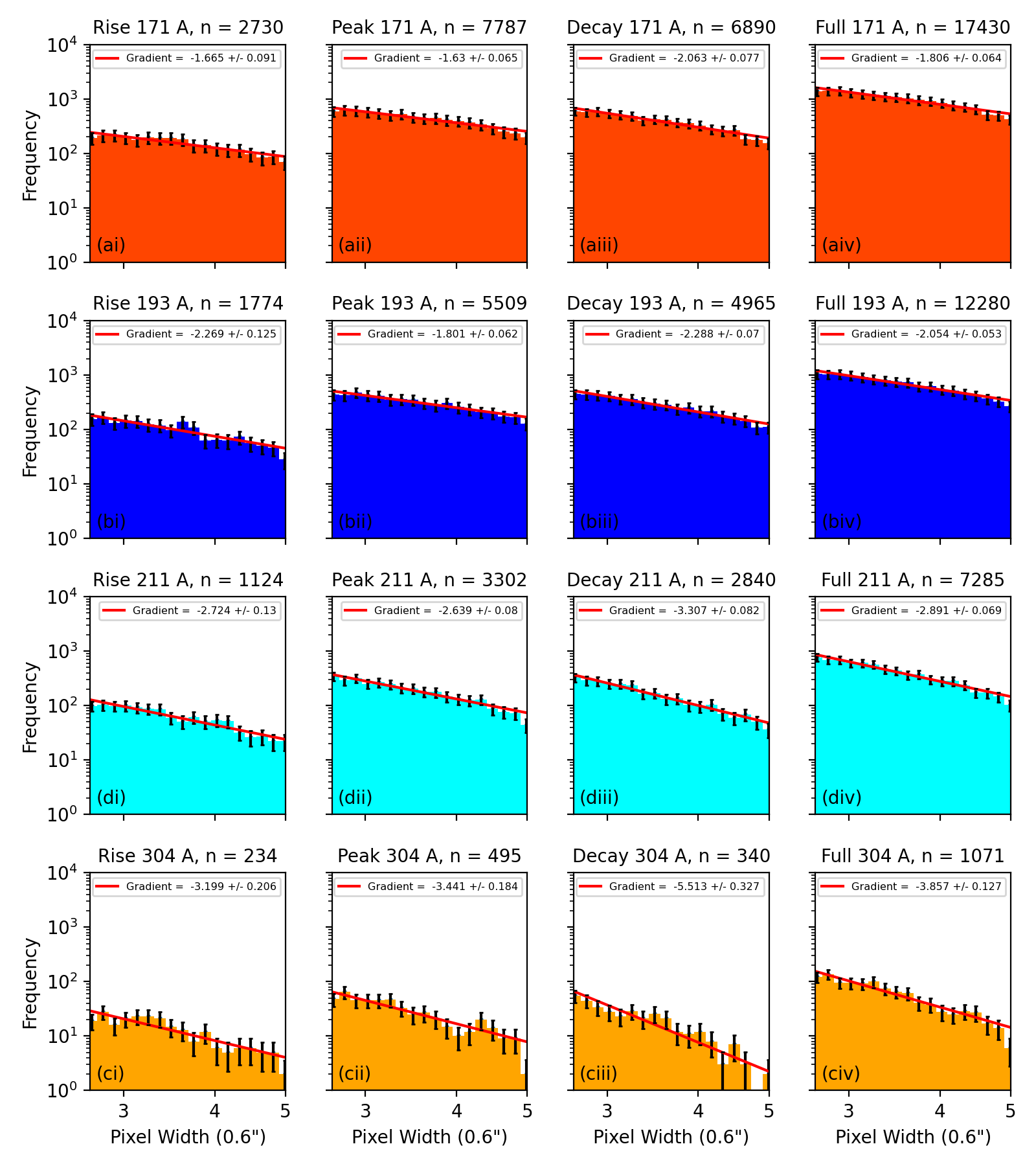}}
    \caption{Width Frequency Diagram with a cutoff from 2.7-5 pixel widths. At this level the differential gradient is closer to the ideal S.O.C case of around 1.5 Variation throughout solar cycle is persistent through different wavelengths but shows more consistency than the wider range.}
    \label{fig:2.7-5ARLoops}
\end{figure*}

More detailed examinations of structure width populations and their power laws can be made by restricting the sample to the range of 2.7 - 5 pixel widths. Hence the physical and observational effects described in Sections 1 and 2 are minimised, and so a more accurate examination of power law gradient slopes can be carried out. Results in this range are presented in Table \ref{table:loops2.7-5} and Fig \ref{fig:2.7-5ARLoops}. The gradients of each distribution are estimated by fitting a power law across the centerpoints of each bin in this range by means of linear regression. This results in a thresholded power law distribution of observed structure widths, indicating a log vs log relationship between width and frequency for structures within a given time period. Subsequently, this relationship is indicative of a probability distribution of structures observed across these time periods. The gradients within different periods across each wavelength and time period within the range of 2.7 - 5 pixels is outlined in Table \ref{table:2.7-5AIAstrucs}.

\begin{table}
\caption{Loop frequency by wavelength and period.}             
\label{table:loopsall}      
\centering                          
\begin{tabular}{c c c c c}        
\hline\hline                 
Wavelength (Å) &Rise & Peak & Decay & Total\\    
\hline                        
   171 &11694&28140&34369&74203 \\      
   193 &8584&23466&27300&59350 \\
   211 &7287&19848&22158&49293\\
   304 &1492&3346&3178&8016\\
\hline                                   
\end{tabular}
\end{table}




\begin{table}
\caption{Loop frequency by wavelength and period. Loop widths between 2.7 and 5 pixels.}             
\label{table:loops2.7-5}      
\centering                          
\begin{tabular}{c c c c c}        
\hline\hline                 
Wavelength (Å) &Rise & Peak & Decay & Total\\    
\hline                        
   171 &2730&7787&6890&17430 \\      
   193 &1774&5509&4965&12280 \\
   211 &1124&3302&2840&7285\\
   304 &234&495&340&1071\\
\hline                                   
\end{tabular}
\end{table}



\begin{table}
\caption{Width frequency power law gradient by time period and wavelength - 2.7-5 pixel widths}             
\label{table:2.7-5AIAstrucs}      
\centering                          
\begin{tabular}{c c c c}        
\hline\hline                 
Wavelength (Å) & Rise & Peak & Decay \\    
\hline                        
   171 & 1.67 $\pm$ 0.09 & 1.63 $\pm$ 0.07  & 2.06 $\pm$ 0.08 \\      
   193 & 2.37 $\pm$ 0.13 & 1.80 $\pm$ 0.06    & 2.29 $\pm$ 0.07 \\
   211 & 2.72 $\pm$ 0.13 & 2.64 $\pm$ 0.08    & 3.31 $\pm$ 0.08 \\
   304 & 3.20 $\pm$ 0.21 & 3.44 $\pm$ 0.18    & 3.86 $\pm$ 0.13 \\
\hline                                   
\end{tabular}
\end{table}


Within the Rise period, gradients of structure populations are generally less steep than those seen in the decay phase, but are equally or more steep than gradients seen within the Peak phase. This can be seen in 171 Å, with the magnitude of the rise gradient being 1.67 $\pm$ 0.09, and the magnitude of the Peak and Decay periods being 1.63 $\pm$ 0.07 and 2.06 $\pm$ 0.08 respectively. This is similarly mirrored in 193 Å and 211 Å, with 304 Å demonstrating a significantly steeper gradient in the Decay phase (-5.513 $\pm$ 0.33) than would be expected from trends seen in other wavelengths. However this is likely due to low numbers (< 100) of high width (> 4 pixels) affecting the slope of the power law. 

Overall, these distributions display a difference in power law gradients between wavelengths for structures captured throughout the entire solar cycle, with gradient magnitudes of 1.81 $\pm$ 0.06, 2.05 $\pm$ 0.05, 2.89 $\pm$ 0.07, and 3.86 $\pm$ 0.13 for 171, 193, 211, and 304 Å structures respectively.

\subsection{Latitude frequency and asymmetry of coronal structures}

\begin{figure*}
    \resizebox{\hsize}{!}
    {\includegraphics[clip]{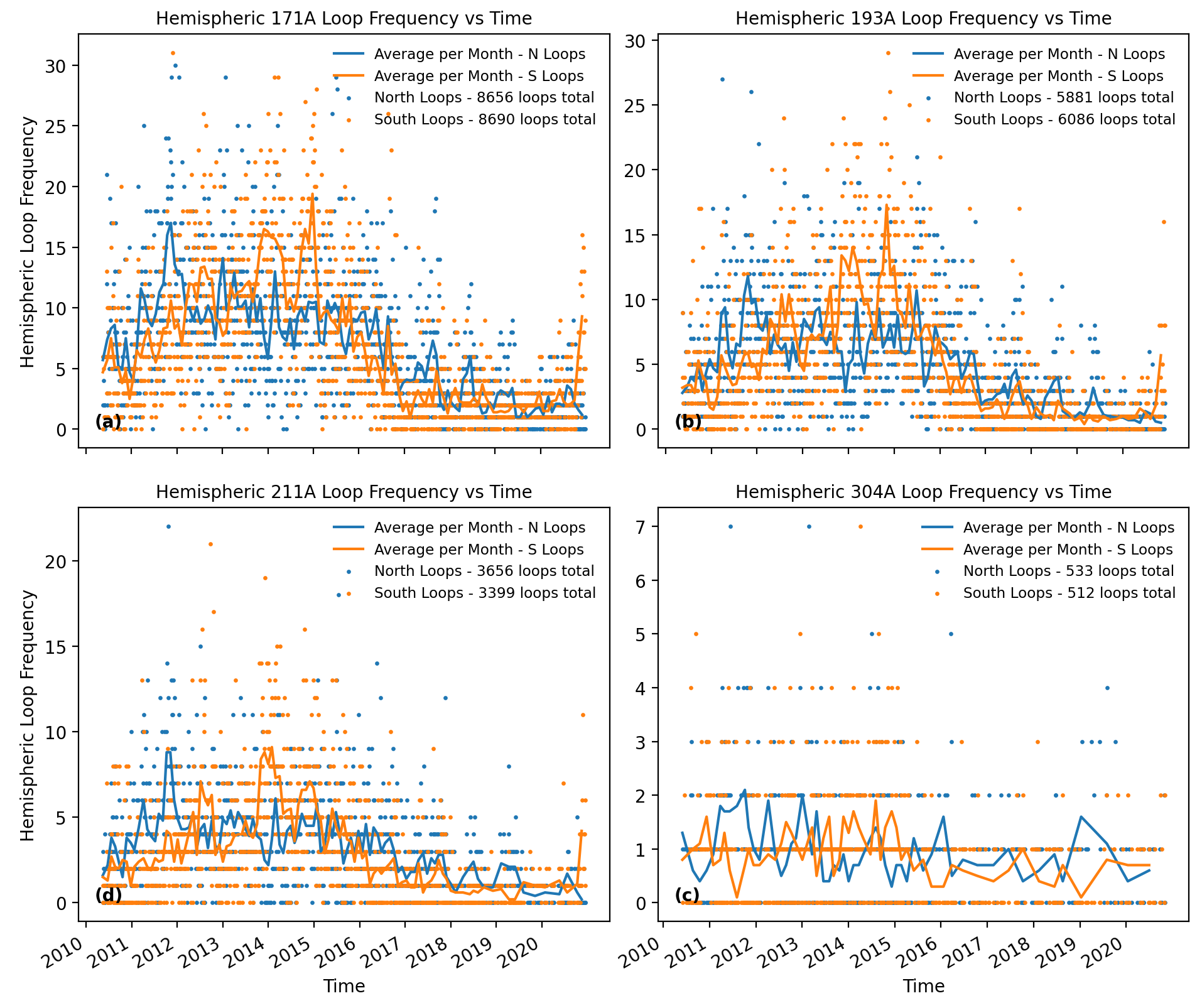}}
    \caption{Hemispheric loop frequency vs time for all wavelengths. Blue indicates northern activity, orange indicates southern activity. Dots and lines indicates loop counts by day and monthly average respectively. All loops are above 2.7 pixels in width.}
    \label{fig:NSLoopsCol}
\end{figure*}

The extent to which coronal structures exhibit any latitudinal asymmetry in different periods of the solar cycle or in different wavelengths could provide insight into the relation between the level of solar magnetic activity and the relative population of coronal structural populations. Possible asymmetry is examined in more detail in Figure \ref{fig:NSLoopsCol}, where the total and average frequencies in the north and south hemispheres for observed coronal widths. Comparing to Figure \ref{fig:NSspots} which examines the total sunspot frequency by hemisphere for Solar Cycle 24, the overall activity of the solar cycle is biphasic. The northern hemisphere (blue) population is greater on average throughout the cycle than the southern hemisphere, but peaks earlier and begins decaying sooner after the start of the "peak" phase. In contrast, the southern hemisphere only peaks in activity towards start of the decay phase, with both hemispheres decaying to negligible activity by the end of cycle 24.



\begin{figure*}
    \resizebox{\hsize}{!}
    {\includegraphics[clip]{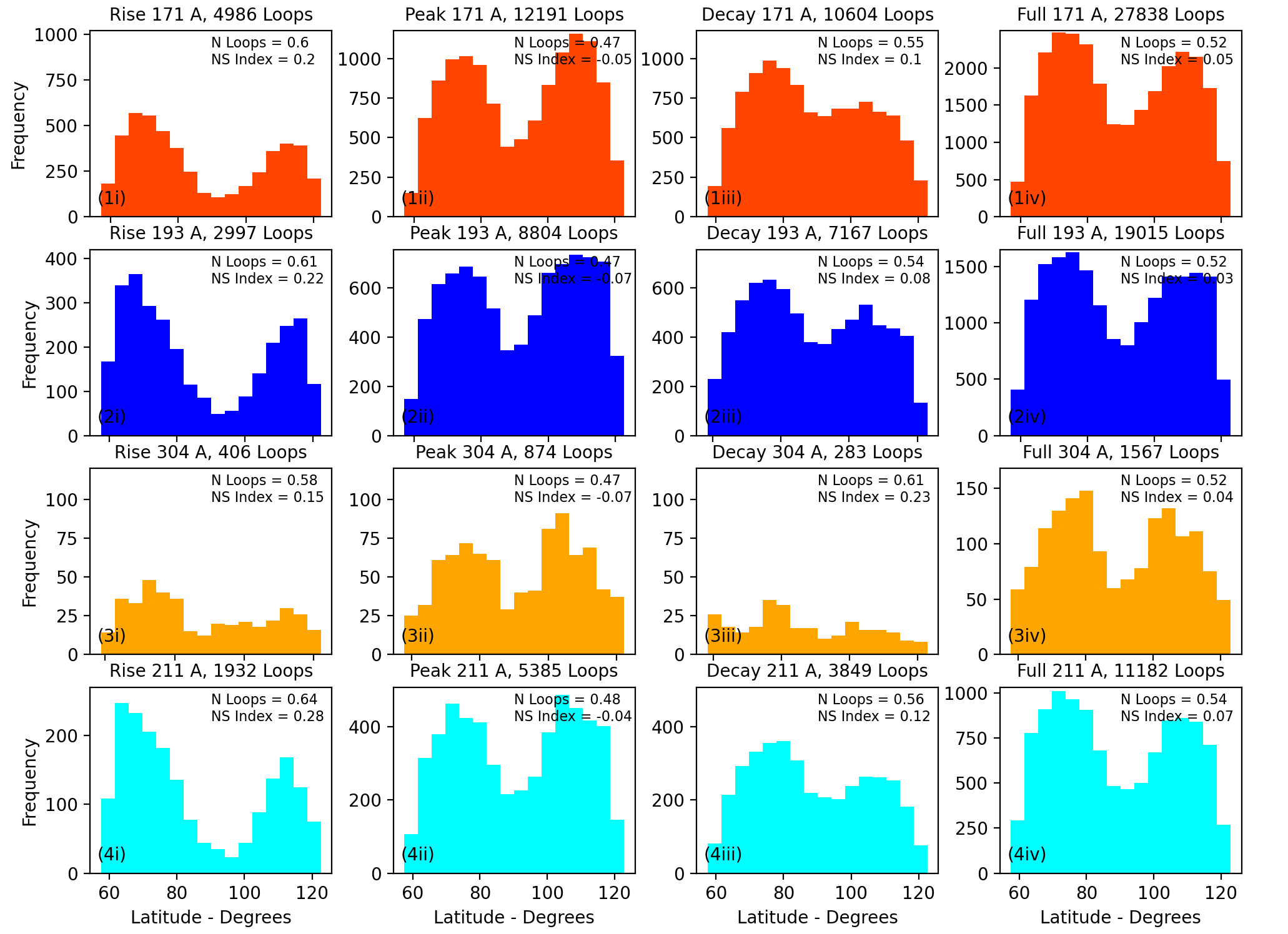}}
    \caption{Latitude frequency diagram for all coronal loops above 2.7 pixel widths by wavelength and period.}
    \label{fig:looplatplot}
\end{figure*}

\begin{table}
\caption{North-South loop asymmetry in three time periods. Left column F represents north hemisphere loops as a fraction of total loops. Right column Å represents N-S loop asymmetry index for chosen period and wavelength.}             
\label{table:NSasymmLoops}      
\centering                          
\begin{tabular}{c c c c c c c c c}        
\hline\hline                 
Wavelength & \multicolumn{2}{c}{Rise} & \multicolumn{2}{c}{Peak} & \multicolumn{2}{c}{Decay} \\  
& F & A & F & A& F & A \\
\hline                        
   Sunspots & 0.68  & 0.36 &  0.37 & -0.27& 0.73 & 0.46 \\ 
   171 Å & 0.59  & 0.18 &  0.46 & -0.07 & 0.51 & 0.03 \\      
   193 Å & 0.6  & 0.19 &  0.45 & -0.09 & 0.5 & 0.00 \\
   211 Å& 0.63  & 0.26 &  0.47 & -0.07 & 0.54 & 0.08 \\
   304 Å& 0.59 & 0.17  & 0.47 &  -0.06 & 0.52 & 0.05 \\
   
\hline                                   
\end{tabular}
\end{table}

Fig \ref{fig:looplatplot} demonstrates the relationship between time period and latitudinal frequency for the population of widths in each wavelength. The fraction of structures which appear in the northern hemisphere is indicated by N. Loops, and the North-South Index (see Eq. \ref{eq:NSind}) given by NS Index for each wavelength. These demonstrate a small but detectable variation from northern dominated structure populations in the Rise phase eg. 0.6 for 171 Å, followed by a switch to a southern dominated population (0.47 for 171 Å) in the Peak phase, and then another change towards northern lead populations in the Decay phase (0.55 for 171 Å). For 171 Å (the most well populated section), this represents a percentage variation of 13 percent, then 8 percent across the two defined phase boundaries. This compares to 14 percent and 7 percent for 193 Å, 16 percent and 8 percent for 211 Å, and 11 percent and 4 percent for 304 Å. These suggest a consistent degree of variation for coronal populations between phases, although the overall changes are small, contributing to population figures which are generally close to parity over the whole Solar Cycle. These figures mirror other indicators of solar activity ie; sunspots as shown in fig \ref{fig:NSspots}. 

It should be noted that the asymmetry arises from the two hemispheres displaying different profiles of activity across the solar cycle, rather than the two cycles simply being totally out of phase or of one being a lesser amplitude than the other. The northern hemisphere appears to to peak earlier in 2012, resulting in a northern lead Rise phase, with the predominantly southern Peak period caused by a more sudden rise of southern lead activity combined with the slowly decaying northern activity.

Taken together, these asymmetries of coronal populations could be explained by a double peaked distribution centered at roughly 70 and 110 degrees (corresponding to the active region belts) from the northern pole, which rise and fall at different times. These values and the profiles of the population demonstrate a visible bifurcation of the overall distribution, with a near parity (0.52 North, 0.48 South) of overall population quantities between hemispheres overall but phases in which one hemisphere is dominant over the other.

The solar magnetic field is expected to be north driven in Solar Cycle 24. However, the results of latitude distribution of active region coronal structures versus time indicate an earlier peaking but extended decaying of coronal activity in the north, and a delayed peaking but accelerated decaying of activity in the south. The highest period of overall sunspot activity occurs due to southern activity rather than that of activity in the hemisphere which dominates the cycle as a whole, hence showing that the phenomenon of hemispheric asymmetry is more nuanced than simply a North vs South predominance per cycle. 

\begin{figure*}
    \resizebox{\hsize}{!}
    {\includegraphics[clip]{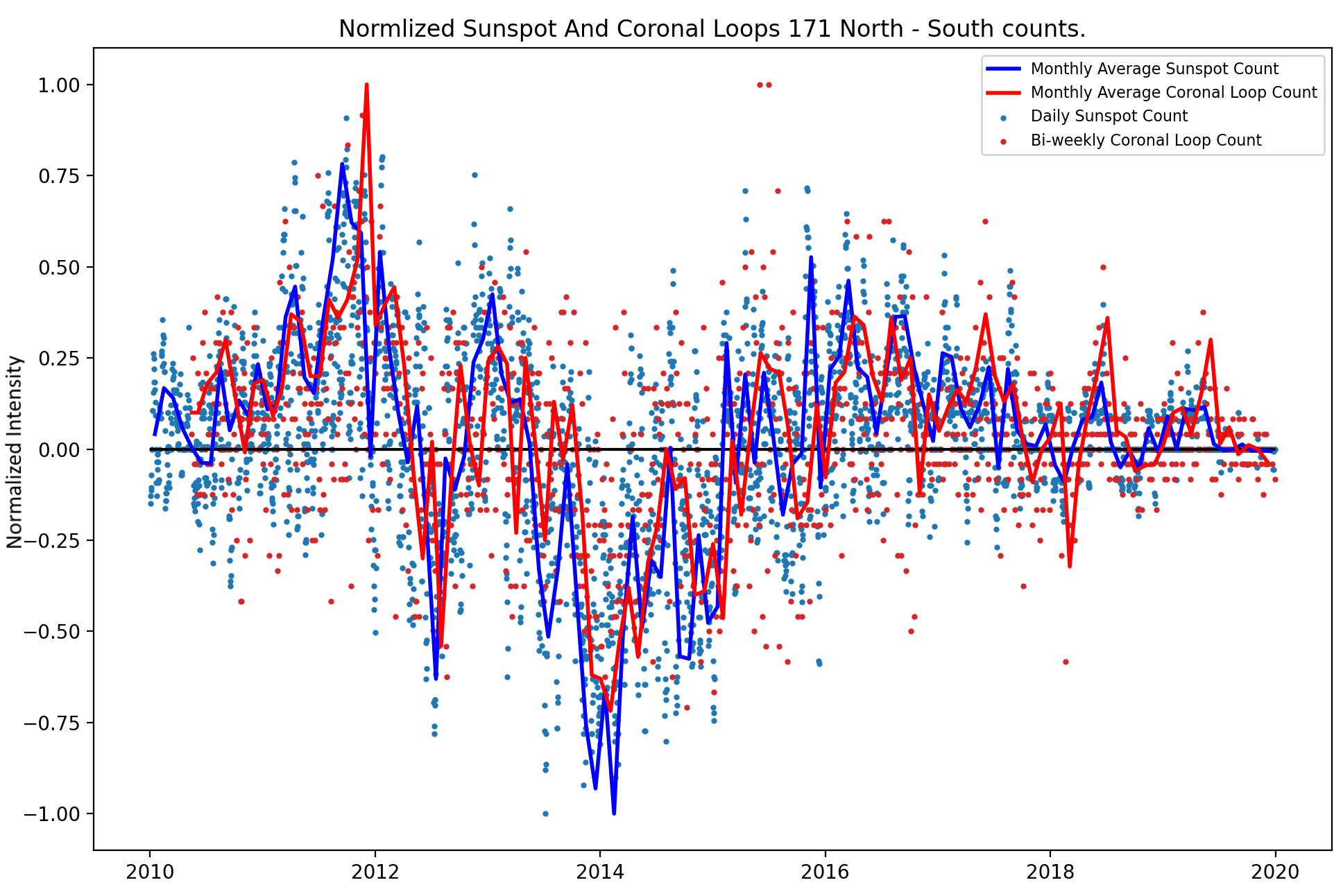}}
    \caption{A normalized frequency diagram of 171 coronal loops and daily sunspot figures vs time over the 24th solar cycle. Lines indicate the monthly averages for both. Sunspot data from SILSO \citep{sidc}}
    \label{fig:Spotsvsloopscounts}
\end{figure*}

A direct comparison of coronal width frequency in north versus south hemispheres to corresponding sunspot numbers yields a strong positive correlation. This is seen in Fig. \ref{fig:Spotsvsloopscounts}, which compares normalized hemispheric sunspot frequencies in raw and smoothed (monthly average) distributions with measured coronal structure equivalents. A lag is seen between some individual peaks of coronal width frequency, but is not consistently before or ahead of corresponding sunspot peaks and likely can be attributed to the difference in where these statistics are collected and (disc versus limb).


There are underlying commonalities between structure populations across different wavelength regimes, indicating that despite being composed of different plasma populations, the underlying conditions required to produce these structures are present. 

A further way to examine these latitudinal variations is to plot a series of Maunder butterfly diagrams \citep{2005MaunderD} of coronal width populations, as seen in Figure \ref{fig:NSLoopsCol} and table \ref{table:fullasymmloops}, using the Royal Belgian Observatory SILSO (Sunspot Index and Long-term Solar Observations) \citep{sidc} daily North and South sunspot coverage figures. 


\begin{table*}
\caption{North-South asymmetry by year for sunspots and coronal structures in different wavelengths. Left column F represents north loops as a fraction of total loops. Right column A represents N-S loop asymmetry index.}             
\label{table:fullasymmloops}      
\centering                          
\begin{tabular}{c c c c c c c c c c c}        
\hline\hline                 
Year & \multicolumn{2}{c}{Sunspots} & \multicolumn{2}{c}{ 171 Å} & \multicolumn{2}{c}{193 Å} & \multicolumn{2}{c}{211 Å} & \multicolumn{2}{c}{304 Å} \\    
& F & A & F & A& F & A& F & A& F & A\\
\hline  
   2010 & 0.651 & 0.303 & 0.588 & 0.177 & 0.55 & 0.1 & 0.554 & 0.107 & 0.454 & -0.092 \\      
   2011&0.687&0.374&0.601&0.202&0.626&0.252&0.664&0.328&0.646&0.291 \\
   2012&0.523&0.045&0.508&0.016&0.481&-0.038&0.509&0.018&0.461&-0.078 \\
   2013&0.408&-0.184&0.5&0.001&0.522&0.043&0.512&0.024&0.493&-0.014 \\
   2014&0.333&-0.333&0.417&-0.165&0.404&-0.193&0.436&-0.127&0.465&-0.07 \\
   2015&0.516&0.032&0.505&0.01&0.486&-0.027&0.493&-0.014&0.625&0.25 \\
   2016&0.741&0.483&0.57&0.139&0.581&0.162&0.628&0.256&0.571&0.143 \\
   2017&0.718&0.436&0.594&0.188&0.595&0.19&0.603&0.207&0.641&0.282 \\
   2018&0.594&0.187&0.548&0.097&0.571&0.141&0.638&0.275&0.4&-0.2 \\
   2019&0.929&0.859&0.591&0.182&0.622&0.244&0.75&0.5&0.826&0.652 \\
\hline                                   
\end{tabular}
\end{table*}

\begin{figure*}
    \centering
    \resizebox{0.95\hsize}{!}
    {\includegraphics[clip]{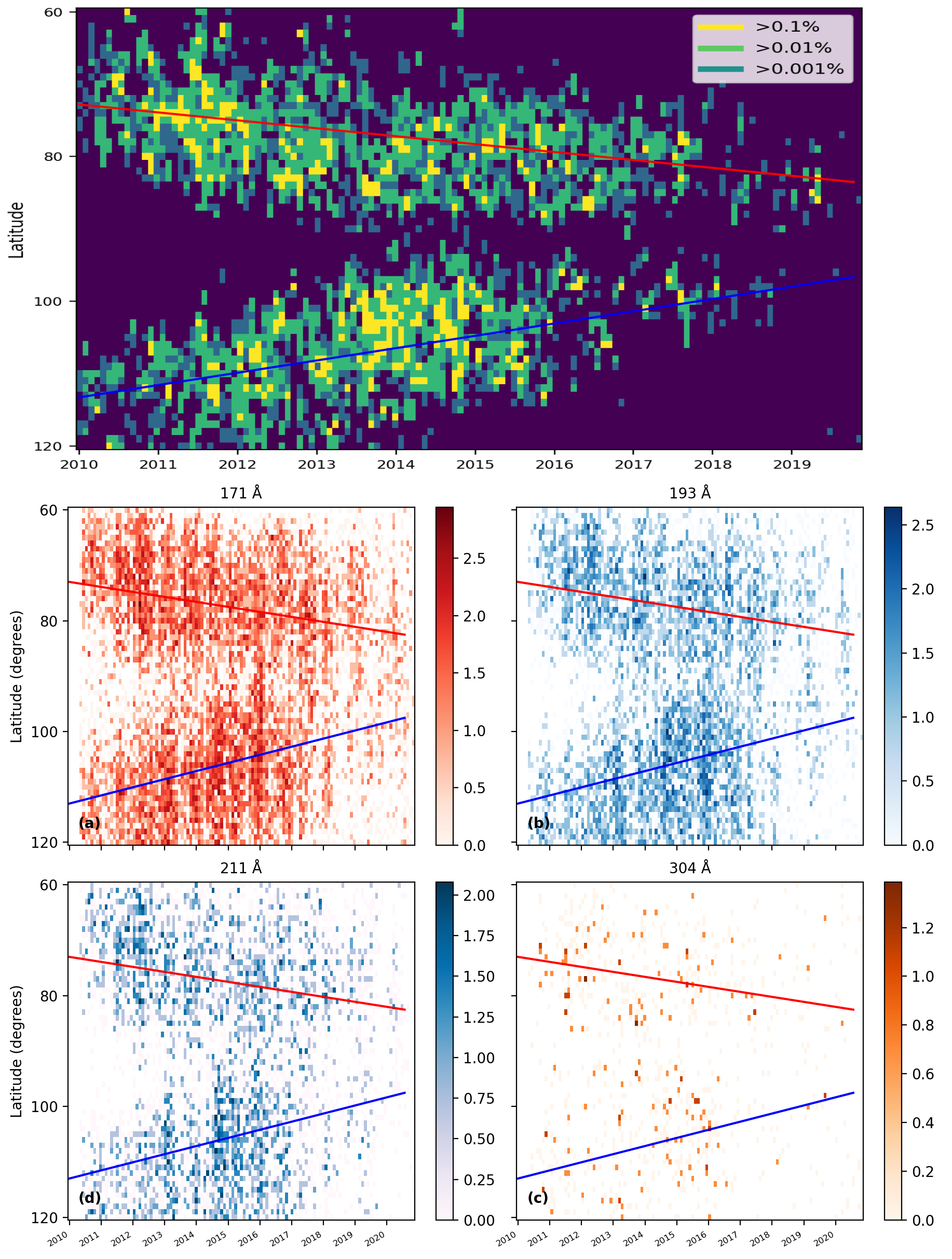}}
    \caption{Diagrams of sunspot umbra area coverage by latitude and time (top)  and coronal width frequency occurence by latitude (bottom). Coronal colour maps correspond to $e^x$ as shown on vertical bar. Line of best fit calculated across yellow pixels (greater than 0.1 percent coverage) in each hemisphere. }
    \label{fig:butterflydiag}
\end{figure*}


A line of best fit was calculated for the hemispheres of the sunspot butterfly diagram (top) by means of liner regression between the cells of highest area coverage (yellow pixels). The gradient and intercept of these lines of best fit for the northern hemisphere (red) is -0.13 and 53, and the southern hemisphere (blue) is  0.08 and 13 respectively. It is seen that coronal widths bear considerable visual similarity to the sunspot distribution - following a similar if slightly broader distributions (40-30 degrees displacement from the equator at the start of the solar cycle, to 15 degrees displacement at the end of the cycle), consistent with Spörer's Law \cite{2014Sporerslaw}. In particular, coronal loops are seen across the whole range of latitudes at all periods, though they are concentrated at positions which correspond to sunspots as would be expected.

Given the intrinsic link between coronal loops and sunspots, it should be expected that coronal loop frequency is strongly correlated to sunspot frequency, this relationship is examined in Figure \ref{fig:Spotsvsloopscounts}. 

\begin{figure}
  \resizebox{\hsize}{!}{\includegraphics{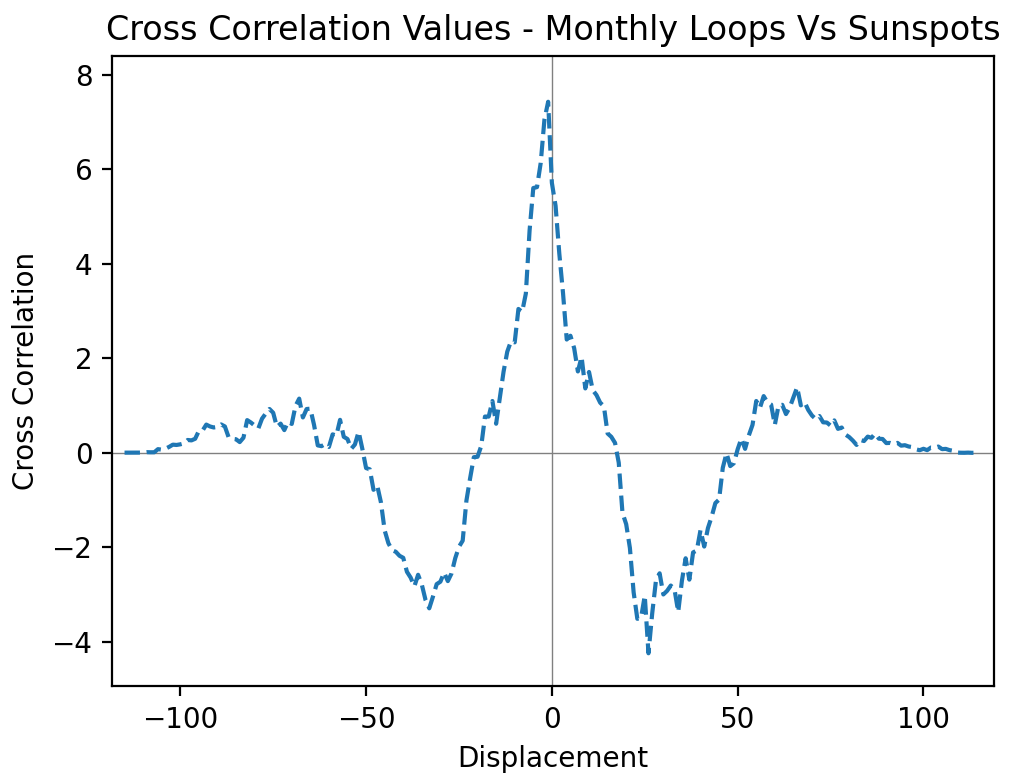}}
  \caption{The cross correlation of monthly average hemispheric sunspot count difference and coronal loop hemispheric frequency difference. The peak value of the cross correlation occurs at position 108, corresponding to the length of the shortest sequence.}
  \label{fig:xcorr}
\end{figure}

To quantify the comparison between the monthly averages of coronal loops and sunspots, a cross correlation was calculated between the overlapping date ranges (1st July 2010 to 30th December 2019), with the standard cross correlation formula;
\begin{equation}
    C_{ab}(\tau) = \sum_{i=1}^{n}a(n + \tau) * \overline{b(n)}
\end{equation}
where C\textsubscript{ab} is the cross correlation of the functions a and b, with time delay $\tau$. 
$\overline{b(n)}$ is the complex conjugate of the function b(n). The results of this cross correlation are displayed in fig \ref{fig:xcorr}. In this case, a is the function of normalized north minus south sunspot frequencies, and b is the function of normalised north minus south loop frequencies for 171 Å populations. This cross correlation is optimal at the 114th displacement value, compared to series length of 116, indicating that the two series are almost perfectly aligned. Therefore, the series of monthly average hemispheric sunspot counts and coronal widths are best aligned without any delay. This is despite some apparent lags seen between spikes of normalized frequency in either series; the lags are either before or after corresponding spikes in sunspot count, though these mostly occur in the decay phase, when there are likely to be low overall counts of both sunspots and coronal structures, making the N-S average more liable to larger variation.

\subsection{Correlation of observed width and latitude}



To quantify any potential correlation, a Spearman rank correlation coefficient (SRCC) and associated p value has been determined for the width versus latitude of each loop in the range of 2-12 pixel widths for each wavelength. The standard equation for SRCC is shown as

\begin{equation}
     R_s = 1 - \frac{6{\sum}d^2_i}{n(n^2-1)}
     \label{eq:SRCC}
 \end{equation}

\begin{table}
\caption{Table demonstrating Spearman Rank Correlation Coefficient results for loop width vs latitude for each wavelength, showing the $R_s$ value, $p$ value, and $n$ number}             
\label{table:widlat}      
\centering                          
\begin{tabular}{c c c c}        
\hline\hline                 
Wavelength (Å) & \textit{R}\textsubscript{s} & \textit{p} & n \\    
\hline                        
   171 & 0.0164 & 0.0063 & 27847 \\      
   193 & -0.0026 & 0.7248 & 19017   \\
   211 & -0.0196 & 0.0378 & 11186 \\
   304 & 0.0450 & 0.07504  & 1567 \\
\hline                                   
\end{tabular}
\end{table}

Here, $\textit{R}$\textsubscript{s} is the correlation coefficient, n is the number of total observations, and \textit{p} is the standard p-value test based on the Student t-distribution. $\sum${d\textsubscript{i}} is the sum of the difference in ranks of each observed loop in width and latitude.

Values of \textit{R}\textunderscore{s} indicate a very low likelihood of any correlation between loop width and position, approaching a perfect lack of correlation to within 0.05 and a large sample size. Values of \textit{p} similarly increases the confidence that the observed lack of correlation is not a coincidence of the data set. The high value of \textit{p} in the 193 Å distribution does not necessarily indicate that the correlation is likely to be caused by an improbable distribution, given the large sample size n and the time taken to observe them. It is therefore unlikely that any correlation might be established between width and latitude based on the information gathered in this dataset in any wavelength.

This indicates a general similarity in the populations of coronal structures imaged in this region regardless of their viewed latitude, and that footpoints at each latitude is equally as likely to produce loops of any coronal width as other latitudes. This could indicate that changes in observed structural width and the required conditions for forming larger loop structures may not vary by latitude as much as by time period. Overall, despite some small variance in structure width across latitude and time period, analysis of latitude versus width indicates that it is a poor indicator of the width of a loop at any time period. At the present time, there is no evidence to suggest any relationship between observed structure width at the limb and latitude of occurrence.

\section{Discussion and Conclusions}

The coronal structures described in this work represent coronal plasma populations at differing temperatures and time periods. It was anticipated that investigating changes in coronal structure widths and latitudes across solar cycle 24 would allow for more detailed insight into S.O.C gradients and asymmetry of the corona over time, and that trends might be identified which had previously escaped detection. This was achieved by comprehensive analysis, using the 171, 193, 211, and 304 Å filters of SDO's AIA instrument. The periods of Rise, Peak, and Decay were defined to describe distinct phases of solar magnetic activity, corresponding to the times between 13/05/2010, 15/11/2011, 15/11/2014, and the end of 2019 respectively.  

Structures were isolated above the limb within a fixed annulus, with segments traced automatically at a rate of one image per three days. Gaussian profiles were fitted to cross sectional intensity values. From these profile measurements, over 50,000 structures were measured in all wavelengths above 2.7 pixel widths (the limit of confident width detection imposed by instrumental constraints). 

From the resulting extensive dataset, analysis of width and latitude was performed. Coronal structure widths were analysed by compiling the width of all loops above 2.7 pixels separated into wavelengths and subsequent populations in the rise, peak, and decay period (as well as an aggregated full time period for comparison). The power law gradient of the produced distribution was compared to both theoretical S.O.C distributions (1.5 for F.D S.O.C) and those measured by other studies of on-limb coronal loops measured in 193 Å (2.7 - 3.1 for \cite{2019Zhiming}, \cite{2017Aschwanden}). These distributions were found to vary across the solar cycle, increasing in magnitude throughout these distinct periods; for example, 1.6 $\pm$ 0.091, 1.63 $\pm$ 0.065, and 2.063 $\pm$ 0.11  for rise, peak, and decay in 171 Å respectively. This trend indicates that observed coronal loop structures appeared to change over time, with less wide structures and more narrow structures present within width distributions as magnetic activity on the sun became less intense. This trend was mirrored across multiple wavelengths, but 171 populations were generally steeper than 193, and 211 were significantly steeper than both 171 and 193 Å. 304 Å populations were much less well populated than other wavelengths, and it was difficult to analyse the gradient of their power law slopes. 

This analysis has revealed two things; (i) that coronal loop width power law distributions change throughout the solar cycle, indicating a possible change to the rate non-linear dissipative events thought to produce coronal loops at their footpoints in the photosphere, and (ii) that these differences are consistent between wavelengths. 

As S.O.C is a statistical model representing ideal cascading exchanges of energy in a uniform grid of arbitrary dimensions, and contains no physics, it is not used to describe the physical processes by which of magnetic, kinetic, and thermal energies are exchanged within the corona, but rather as an approximation of the distributions and statistics that such a system may result in. The F.D estimation of -1.5 for the value of the power law  gradient of the distribution is based upon the scale invariability of the distribution of the size of events across many orders of magnitude, occurring with perfectly efficient exchanges of energy. That some of the observed distributions follow power law profiles with gradient values approaching -1.5 is evidence that this statistical principle can apply to the distribution of coronal structural widths, but discrepancy is to be expected. These discrepancies can be categorised as either observational or physical. Of the two, physical discrepancies are useful for providing information to the physics within the structure forming regions, and observational discrepancies must be controlled for when possible. 

Observational discrepancies are caused by limitations of the equipment such as CCD charge spreading, this causes the intensity "peak" seen in the overall distribution seen in Fig. \ref{fig:WidthsAll} at two pixels, and may also result in steeper than expected gradients as narrower structures are misidentified as wider structures in lower pixel ranges. Another observational complication is line of sight effects such as structure conflation, by which multiple structures overlap within the line of sight and are subsequently misidentified, and can cause thresholding effects (see Eq. \ref{eq:wspreading}. Additionally, background subtraction can result in wider and dimmer structures as being detected as narrower than they are in reality. As there is generally little correlation between coronal structural width and observed intensity \citep{2020Klim}, this can potentially curtail distributions at the wider ranges (>4 pixels). These effects can be mitigated by making repeated observations, and comparing results seen in various periods of time. This allows for observational effects to remain consistent in all measurements, meaning that any changes in the resulting gradients are more descriptive of discrepancies caused by physical processes than aforementioned observational effects.

Physical variations could be caused by effects analogous to those which are known to affect S.O.C-like populations, these include but are not limited to, limited size scaling effects which curtail the size distribution available to cascade over, quenching effects which removes energy in a way which does not contribute to further cascading, variations of dissipation time scales which alter the amount of time available for energy to dissipate throughout the system. Additionally, other assumptions of S.O.C like stochastic addition and that the local or global critical threshold is constant may not be accurate representations of structure forming regions in all cases. Variation in observed parameters (such as width distributions) across solar cycle could in part aid in describing the fundamental effects such as reconnection rates and mechanisms which influence the formation of coronal structure, as differing gradients will necessitate differing frequencies of reconnection events which generate observable structural widths, which must be consistent with predicted values. Observed variations from ideal FD-SOC could indicate that physical processes responsible for the appearance of coronal structures may be less similar to the ideal assumptions in later periods of the solar cycle than in the earlier, more active periods. The value of the power law slope of the frequency distribution of loop widths D(w) is expected to be analogous to the same power law gradient of cascading events $\alpha$\textsubscript{s} taking place within the emerging region assuming that principles of S.O.C apply to the phenomenon of coronal loops. 

Similarly to coronal widths, coronal loop latitudes were recorded and compiled. These latitudes were similarly divided into rise, peak, decay populations and the distributions examined. (See fig. \ref{fig:looplatplot}) These distributions revealed asymmetries in coronal loop populations in north and south hemispheres which varied throughout the solar cycle, but which remained relatively consistent between wavelengths; Northern loop fraction of 0.5 for 171 Å, 0.49 for 193 Å, and 0.54 for 211 Å,0.521 for 304 Å. While these fractions are similar across each wavelength, there is significant variation within the solar cycle, with periods of southern lead activity reaching as high as 60 percent in 171 Å. 

Following this, analysis of the relation between coronal loop width and latitudinal position was performed within each of the periods of study mentioned previously. The purpose of this analysis was to examine the relation between coronal activity and latitude. This is examined statistically by means of a test of correlation coefficient of width and latitude for each loop recorded above 2 pixels. The results (table \ref{table:widlat}) show an R value magnitude of less than 0.05 for every wavelength. This indicates that no meaningful relation can be determined between loop latitude and width independent of time period, and that the conditions at each latitude do not vary substantially with regards to coronal loop formation.  

\subsection{Future Work}

More detailed examination of loop populations present in existing and future datasets, such as examination of historical datasets within the SOHO EIT, and Solar Cycle 25 with the ongoing SDO AIA and targeted studies of active regions with Solar Orbiter's EUI. This will be of use in understanding how the corona and its magnetic framework changes and can change over time. Additionally, more work regarding the exact change of these gradients and how they may relate to heating mechanisms within/without active regions and periods of the solar cycle may be useful as a probe of exact heating mechanisms and thresholds. The resulting distributions of populations of coronal structures are caused by specific physical processes and conditions within the structure forming regions. Due to the stochastic nature of their formation, however, these parameters may only be detectable when analysing the statistical profiles of many tens of thousands of these structures. It is hoped that further understanding and measurement of these profiles could aid in determining the likely predominance of heating profiles and their action within the corona throughout the solar cycle, by limiting the possible heating scenarios to those which could be capable of producing the matched distributions in these large samples.



\section*{Acknowledgements}

This research was funded by the University of Central Lancashire as part of the Jeremiah Horrocks Institute Postgradute Research program fund. utilizing data made available by NASA, the ESA, and SILSO. 


\section*{Data Availability}

The data underlying this article and the software used is freely accessible from the UCLAN Data Archive, at \url{https://doi.org/10.17030/uclan.data.00000377}.



\bibliographystyle{mnras}
\bibliography{main} 








\bsp	
\label{lastpage}
\end{document}